\documentclass{jfm}
\usepackage[utf8]{inputenc}
\usepackage{color}
\usepackage{amsmath,amsfonts}
\usepackage{natbib}
\usepackage{gnuplottex}
\usepackage{overpic}
\usepackage{subcaption}

\title{Variational methods for finding periodic orbits in the incompressible Navier-Stokes equations}
\author{J. P. Parker\aff{1}
  \corresp{\email{jeremy.parker@epfl.ch}}, \and T. M. Schneider\aff{1}}

\affiliation{\aff{1} Emergent Complexity in Physical Systems Laboratory (ECPS), \'Ecole Polytechnique F\'ed\'erale de Lausanne, CH-1015 Lausanne, Switzerland}

\begin{document}

\maketitle

\begin{abstract}
Unstable periodic orbits are believed to underpin the dynamics of turbulence, but by their nature are hard to find computationally.
We present a family of methods to converge such unstable periodic orbits for the incompressible Navier-Stokes equations, based on variations of an integral objective functional, and using traditional gradient-based optimisation strategies. 
Different approaches for handling the incompressibility condition are considered.
The variational methods are applied to the specific case of periodic, two-dimensional Kolmogorov flow and compared against existing Newton iteration-based shooting methods. While computationally slow, our methods converge from very inaccurate initial guesses.
\end{abstract}

\section{Introduction}

Exact coherent structures, and particularly unstable periodic orbits, are believed to underpin the dynamics of turbulence, by acting as a `backbone' of the chaos-supporting set in phase space \citep{kawahara2001periodic,kawahara2012significance,ChaosBook}.
The most successful method for finding periodic orbits has been the shooting method based on Newton iteration, which, though fast and easy to implement on top of existing time-marching codes, is notable for a relatively small radius of convergence, requiring extremely accurate initial guesses. Extensions to Newton's method, employing a hookstep procedure, have been successful in enlarging this radius of convergence \citep{viswanath2007recurrent, schneider2010snakes,chandler2013invariant, dijkstra2014numerical, vanveen2019periodic, tuckerman2020computational}, but the issue of accurate guesses remains.
Alternative methods for finding exact coherent states have been proposed, for example \citet{lucas2021stabilisation} applied the idea of time-delayed feedback to stabilise otherwise unstable steady states and travelling waves.

\citet{lan2004variational} developed an alternative to the shooting method which continuously deforms temporal loops in phase space into periodic orbits. The method was successfully applied to the Kuramoto-Sivshinsky equation \citep{lan2008unstable}, but it does not scale well to larger systems because of the need for explicit Jacobian matrices.
\citet{boghosian2011new} applied this method to a Lattice-Boltzman formulation of Navier-Stokes, but noted the prohibitively large computational requirements.
\citet{azimi2020adjoint} used a matrix-free method inspired by \citet{farazmand2016adjoint} to find periodic orbits in the Kuramoto-Sivashinsky equation.
Though they successfully demonstrated the method, challenges remain towards the ultimate goal of applying this method to find unstable periodic orbits in fluid dynamics, as the studied one dimensional equation differs from the two- and three-dimensional incompressible Navier-Stokes equations in important aspects.  
In particular, the divergence-free condition and the presence of the pressure field, satisfying a Poisson equation with velocity-dependent source-term, render the problem non-local. These non-local effects due to incompressibility are absent in simpler, local, partial differential equations but require careful consideration when formulating variational methods for computing periodic orbits of the incompressible Navier-Stokes equations.

As a stepping-stone towards full three dimensional turbulence, several previous authors \citep{chandler2013invariant,lucas2015recurrent,lucas2021stabilisation} have studied the particular case of periodic, two-dimensional Kolmogorov flow, at relatively low Reynolds number.
This has several attractions:
in two dimensions the number of degrees of freedom in the system is greatly reduced, yet this flow supports turbulence-like chaos;
the lack of physical boundaries simplifies  computations, allowing the use of a Fourier (pseudo-)spectral discretisation;
and non-trivial invariant solutions, which have been studied in detail for decades \citep{meshalkin1961investigation,obukhov1983kolmogorov,platt1991investigation,fylladitakis2018kolmogorov} appear at Reynolds numbers achievable with the use of relatively low spatial resolutions. 
It is also possible to achieve similar two-dimensional flows in experiments. \citet{suri2020} found 7 numerically converged periodic orbits and showed that these closely matched the dynamics in an experimental configuration of Kolmogorov flow. Recently, \citet{yalniz2021} used an existing Newton-hookstep method to find 18 periodic orbits in \emph{three}-dimensional Kolmogorov flow, which was sufficient to reduce the dynamics of the system to a remarkably simple Markov chain model.
We also concentrate on the case of 2D Kolmogorov flow, for the reasons listed above, as well as for comparison with this previous work.

In this paper, we study methods which follow ideas of \citet{azimi2020adjoint}, but which are formulated for 2D Kolmogorov flow and explicitly address the incompressibility constraints of Navier-Stokes.
In section \ref{sec:objectives} we present the governing equations of the flow, formulate objective functionals that are minimised at periodic orbits, and derive the necessary gradients of these.
Formulations are presented which tackle incompressibility in different ways: either it is implicitly enforced by the governing equations; it is included in the objective functional; or we explicitly project onto and restrict to the incompressible subspace.
In section \ref{sec:optimisation}, we discuss possible optimisation strategies for these objective functionals.
Section \ref{sec:comparison} compares three variational methods for two test cases and then applies the most successful to a larger number of possible periodic orbits, comparing against the traditional shooting method.
Concluding remarks  are given in section \ref{sec:conclusion}.

\section{Objective functionals for unstable periodic orbits}
\label{sec:objectives}

The well-known incompressible Navier-Stokes equations, in non-dimensional form, describe the evolution of the 2D flow velocity $(u,v)$ with time $t$
\begin{align}
    \partial_t u + u \partial_x u + v \partial_y u &= - \partial_x p + \frac{1}{Re} \Delta u + f_x,\label{eq:governing_PV1}\\
    \partial_t v + u \partial_x v + v \partial_y v &= - \partial_y p + \frac{1}{Re} \Delta v + f_y,\label{eq:governing_PV2}
\end{align}
subject to the incompressibility constraint
\begin{equation}
    \partial_x u + \partial_y v = 0,\label{eq:governing_incompressibility}
\end{equation}
where the pressure $p$ acts as a Lagrange multiplier to enforce the incompressibility.
Here $Re$ is the usual Reynolds number, obtained by non-dimensionalising the equation with respect to the forcing amplitude (note that different definitions of $Re$ are used for Kolmogorov flow in the literature), and $f_x$ and $f_y$ are the x- and y-components of a body force.
Henceforth, following previous authors \citep{chandler2013invariant, lucas2021stabilisation}, we will take $f_x=\sin{4y}$, $f_y=0$, and consider a doubly periodic box of size $2\pi \times 2\pi$. Throughout, we take $Re=40$, which is sufficiently large to allow extensive chaotic behaviour, but permits the use of relatively low resolutions to expedite computations. 
The system (\ref{eq:governing_PV1}-\ref{eq:governing_incompressibility}) with this choice of forcing exhibits several symmetries which permit special exact coherent structures. Continuous invariance under the family of translations $x\mapsto x+\phi$ for $\phi\in\mathbb{R}$ allows travelling waves and relative periodic orbits with a drift velocity $c$ in this direction. The system is also invariant under the discrete transformation $(x,y)\mapsto (-x,y+\pi/4)$, which permits relative periodic orbits, though these are equivalent to stationary periodic orbits with eight times the period.

By differentiating (\ref{eq:governing_PV1}) and (\ref{eq:governing_PV2}), we can derive a governing equation for the vorticity $\omega = \partial_x v - \partial_y u$,
\begin{equation}
    \partial_t \omega +\partial_y\psi\partial_x\omega -\partial_x \psi\partial_y\omega = \frac{1}{Re}\Delta\omega +4\cos{4y}, \label{eq:governing_SV1}
\end{equation}
where the streamfunction $\psi$ is defined such that $u=\partial_y\psi$ and $v=-\partial_x\psi$ and so $    \omega = - \Delta \psi.$

These two formulations, called respectively primitive variables (PV) and streamfunction-vorticity (SV), are equivalent in two dimensions. 
The latter uses only two variables rather than three, since incompressibility is directly built into the formulation, but this advantage is lost when the equations are extended to three spatial dimensions. It is further possible to eliminate $\omega$ to give one equation solely in terms of $\psi$. As variational methods based on the resulting fourth order equation were found to be very inefficient, we do not discuss this formulation further.

Care must be taken when comparing the boundary conditions between the two formulations. 
Assuming a doubly periodic streamfunction in the streamfunction-vorticity formulation implies that $u$ and $v$ have no mean component over the domain, i.e. no net flow, which is not a priori imposed by the primitive variables formulation with periodic boundary conditions on $u$ and $v$.
However, it is straightforward to show that (\ref{eq:governing_PV1}-\ref{eq:governing_incompressibility}) conserve the net flow,
which is implicitly specified by the choice of initial conditions.
Therefore, the system modelled by the streamfunction-vorticity formulation represents an invariant subspace of the primitive variables system.
For initial conditions with no net flow both formulations are equivalent, and we henceforth make this choice.

Let us consider closed `loops' in the phase space of the system, where each component ($u$, $v$ and $p$ or $\psi$ and $\omega$) is defined analogously to
\begin{equation*}
u(x,y,s):[0,2\pi)\times[0,2\pi)\times[0,2\pi)\to\mathbb{R},
\end{equation*}
periodic in all three dimensions, and assumed to be sufficiently smooth. Similarly we can define loops of $v$, $p$, $\psi$ and $\omega$.
Introducing a period $T>0$, such loops are exact periodic orbit solutions of the system if they satisfy the governing equations with $\partial_t = \frac{2\pi}{T}\partial_s$.
In this perspective, $s$, which parameterises the loop in phase space, is a scaled time variable.
From the governing equations of the two different formulations, we can define objective functionals based on the PV formulation
\begin{multline}
        J_{PV}[u,v,p,T] = \frac{1}{2}\int  \Bigg\{
        \left(\frac{2\pi}{T}\partial_s u + u\partial_x u +v\partial_y u +\partial_x p -\frac{1}{Re}\Delta u - \sin{4y}\right)^2\\
        +\left(\frac{2\pi}{T}\partial_s v + u\partial_x v +v\partial_y v +\partial_y p -\frac{1}{Re}\Delta v\right)^2
        +\left(\partial_x u + \partial_y v\right)^2
        \Bigg\}\mathrm{d}V    ,
\end{multline}
and for the SV formulation
\begin{multline}
        J_{SV}[\psi,\omega,T] = \frac{1}{2}\int  \Bigg\{
        \left(\frac{2\pi}{T}\partial_s \omega + \partial_y\psi\partial_x\omega -\partial_x \psi\partial_y\omega - \frac{1}{Re}\Delta\omega -4\cos{4y}\right)^2\\
        +\left(\omega+\Delta\psi\right)^2
        \Bigg\}\mathrm{d}V    ,
\end{multline}
where the integrals are taken over $(x,y,s)\in[0,2\pi)^3$.
Both functionals are necessarily greater than or equal to zero, with equality if and only if the governing equations are satisfied everywhere on the loop. Therefore, finding zeros of these functionals gives us, in general, periodic orbits of the system. 
Equilibria are a special case, for which $\partial_s u = 0$ etc., and the value of $T$ is arbitrary.
While there is an exact correspondence between zeros, i.e. global minima, of both objective functionals, local minima do not generally translate from one formulation to the other.

To minimise these objective functionals, we derive gradients. For the scalar variable $T$, these are simple partial derivatives, for example
\begin{equation*}
    \frac{\partial J_{SV}}{\partial T} =
    \int 
        -\frac{2\pi}{T^2}\partial_s \omega\left(\frac{2\pi}{T}\partial_s \omega + \partial_y\psi\partial_x\omega -\partial_x \psi\partial_y\omega - \frac{1}{Re}\Delta\omega -4\cos{4y}\right)
        \mathrm{d}V.
\end{equation*}
For the loop variables $u$, $v$, $p$, $\psi$ and $\omega$, we must instead derive variational derivatives which are defined at all $(x,y,s)\in[0,2\pi)^3$, such as
\begin{equation*}
    \frac{\delta J_{SV}}{\delta \omega} = -\frac{2\pi}{T}\partial_s I - \partial_x(I\partial_y \psi)+ \partial_y(I\partial_x \psi) - \frac{1}{Re}\Delta I + (\omega+\Delta\psi),
\end{equation*}
where for clarity we have defined
\begin{equation*}
    I = \frac{2\pi}{T}\partial_s \omega + \partial_y\psi\partial_x\omega -\partial_x \psi\partial_y\omega - \frac{1}{Re}\Delta\omega -4\cos{4y}.
\end{equation*}
Since we have no physical boundaries in this problem, such variational derivatives are relatively straightforward to derive.

\subsection{Relative periodic orbits}
As mentioned above, there are symmetries in both the $x$ and $y$ directions of the system, which permit the existence of relative periodic orbits (RPOs). We choose not to study RPOs which exploit the discrete symmetry in the $y$ direction, though this would be a simple extension of our method, where necessary considering two periods of the orbit to eliminate the discontinuity caused by the change of sign in the $x$ direction, as in \citet{chandler2013invariant}.

For an RPO with a drift velocity $c$ in the $x$ direction relative to the frame in which there is no net flow, we can transform into a frame moving with the RPO, so that it becomes a simple periodic orbit.
This is achieved by replacing $\frac{2\pi}{T}\partial_s u$ by $\frac{2\pi}{T}\partial_s u-c\partial_x u$ etc., so that extra terms appear in the gradients. We also now must optimise with respect to $c$, using the partial derivative
\begin{equation*}
    \frac{\partial J_{SV}}{\partial c} =
    \int - \partial_x \omega\left(\frac{2\pi}{T}\partial_s \omega-c\partial_x\omega + \partial_y\psi\partial_x\omega -\partial_x \psi\partial_y\omega - \frac{1}{Re}\Delta\omega -4\cos{4y}\right)\,\mathrm{d}V
\end{equation*}
and similarly for $J_{PV}$.
Since a periodic orbit is a special case of an RPO with $c=0$, and travelling waves and equilibria are also captured, we use these extended objective functionals for all computations.
The full objective functionals and all the gradients are given in the supplementary materials.
It is important to note that the gradients for the PV formulation have no mean component, assuming that $u$ and $v$ do not, and so the (zero) net flow is conserved along gradients.

With the inclusion of the phase speed drift velocity $c$, the space of loops in each of the formulations is represented as a vector space of tuples
\begin{equation*}
    X = (u,v,p,T,c) \quad\mathrm{or}\quad X = (\omega,\psi, T,c),
\end{equation*}
with inner products defined on these in the obvious way:
\begin{align*}
    \left<X_1,X_2\right>_{PV} &= \int \left(u_1u_2+v_1v_2+p_1p_2\right)\mathrm{d}V + T_1T_2 + c_1c_2,\\
    \left<X_1,X_2\right>_{SV} &= \int \left(\omega_1\omega_2 + \psi_1\psi_2\right)\mathrm{d}V + T_1T_2 + c_1c_2,
\end{align*}

\section{Optimisation methods for minimizing the objective functionals}
\label{sec:optimisation}
Minimising the value of the objective functionals we have defined, which are single valued functions of (at least in a discretised sense) a very large number of unknown variables with known derivatives, is best performed through gradient-based methods.
This contrasts with methods by which we would directly find zeros of the integrands through Newton iteration and related methods, which require the same number of equations as unknowns.
Descent methods find only local minima of the function, but to find exact solutions of the governing equations, we need to find zeros, i.e. global minima, of these non-negative functionals.
The methods we present give monotonically decreasing and/or globally convergent results, but this does not mean we are guaranteed to converge on exact solutions to the governing equations.

The simplest method is gradient descent with a fixed step size $\epsilon$ at each iteration. Previous authors \citep{farazmand2016adjoint, azimi2020adjoint} have considered a dynamical system by the introduction of a fictitious time $\tau$ such that 
\begin{equation}
    \label{eq:fictitious}
    \frac{d u}{d \tau} = -\frac{\delta J}{\delta u}.
\end{equation}
Simple gradient descent is equivalent to discretising this system with a forward-Euler scheme, and some improvements may be found by considering terms implicitly or using higher order schemes, or adaptive timesteppers.
This differs from the method of \citep{lan2004variational}, whose fictitious system evolves in a direction equivalent to an infinitessimal step of a Newton-Raphson method, as opposed to in the steepest descent direction for the cost function. This offers much faster convergence, but requires the computation of an explicit Jacobian matrix.

However, we are not interested in solving the fictitious system accurately, but merely in finding its stable fixed points.
Faster convergence is found using the nonlinear conjugate-gradient method \citep{hager2006survey}, which requires only the first order gradients. If the Hessian matrix were also calculated, Newton's method could be employed.
We implemented this with the \citet{fletcher1964function} choice of the conjugate-gradient parameter $\beta$.
The optimal step size $\alpha$ was chosen by a line search, from an initial guess of $\alpha=10^{-5}$ for $J_{PV}$ and $\alpha=10^{-7}$ for $J_{SV}$.
The strong \citet{wolfe1969convergence} conditions refine this initial guess, using the tunable parameters $0<c_1<c_2<1$, where a smaller $c_2$ enforces a sufficiently large step that the objective functional decreases rapidly, while a larger $c_1$ enforces that we do not overshoot and take too large a step.
Through trial and error, we found acceptable values for these parameters to be $c_1=10^{-5}$ and $c_2=0.999$, though we make no claims that these are optimal.
With this choice of parameters, typically only one iteration one the line search algorithm was necessary before the Wolfe conditions were satisfied.
The significantly smaller step sizes required for the conditions to be satisfied for SV mean that this method takes much longer to converge, as discussed in section \ref{sec:comparison}.

Since $x$, $y$ and $s$ are all periodic, the fields can be easily expressed as Fourier series, using a pseudo-spectral approach -- with 2/3 dealiasing -- for the nonlinear terms.
A new code was developed for this work in C++, with OpenMP parallelisation. Following \citet{chandler2013invariant}, we perform time integrations using Heun's method, with Crank-Nicolson on the viscous terms at every substep.
We concentrate on the specific case $Re=40$, which is sufficiently high to allow fully chaotic behaviour, but low enough so that a spatial resolution of $64\times64$ gridpoints gives accurate results when compared with previous work.
A resolution of $64$ gridpoints was also used in the temporal dimension, which was found to be more than sufficient for even the longest periodic orbits studied. A $2/3$ dealiasing rule was applied in the spatial but not temporal dimensions.

\subsection{Leray projection}

In section \ref{sec:objectives} we gave objective functionals for two different formulations of the system.
The PV formulation does not assume the velocity field is divergence-free, and instead includes the incompressibility condition within the objective functional, whereas the alternative SV formulation automatically enforces incompressibility.
Since we know that any converged solution must be divergence free, it may also be desirable to enforce this during the optimisation within the PV formulation. However, care must be taken to ensure that the resulting method still ensures the monotonic decrease of the objective functional.
For this, we recall the two-dimensional Leray projection operator \citep{temam2001navier}
\begin{equation*}
    \mathbf{P}:(u,v)\mapsto(u,v)-(\partial_x,\partial_y)\Delta^{-1}(\partial_x u + \partial_y v),
\end{equation*}
where $\Delta^{-1}$ is the inverse Laplacian, here taken with periodic boundary conditions.
For any $(u,v)$, observe that $\mathbf{P}(u,v)$ is divergence-free.
Then for the usual inner product,
\begin{align*}
    \left<\mathbf{P}(u,v),(u,v)\right>&=\left<\mathbf{P}(u,v),\mathbf{P}(u,v)\right>+\left<\mathbf{P}(u,v),\nabla\Delta^{-1}(\partial_x y + \partial_y v)\right>\\
    &=\left\|\mathbf{P}(u,v)\right\|^2-\left<\nabla\cdot \mathbf{P}(u,v),\Delta^{-1}(\partial_x u + \partial_y v)\right>\\
    &=\left\|\mathbf{P}(u,v)\right\|^2 \geq 0,
\end{align*}
where we have used the fact that $\nabla$ is adjoint to $-\nabla\cdot$ with respect to the inner product.
Therefore, using an update direction of $-\mathbf{P}\left(\frac{\delta J_{PV}}{\delta u},\frac{\delta J_{PV}}{\delta v}\right)$ instead of $-\left(\frac{\delta J_{PV}}{\delta u},\frac{\delta J_{PV}}{\delta v}\right)$ will still converge to a minimum of the objective functional,
since the projected gradient is never directed opposite to the unconstrained gradient.

This leaves us with three potential variational methods: optimising $J_{SV}$ (SV), optimising $J_{PV}$ (PV), or optimising $J_{PV}$ with Leray projection of the gradient at each step to enforce the divergence-free condition (PV-LP).

\section{Comparison of methods}
\label{sec:comparison}

\begin{figure}
        \centering
\includegraphics[width=0.4\textwidth]{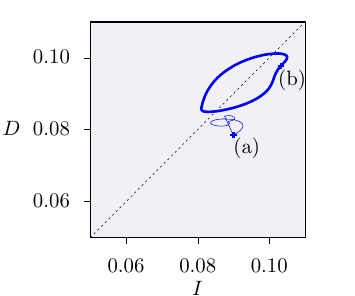} \includegraphics[width=0.4\textwidth]{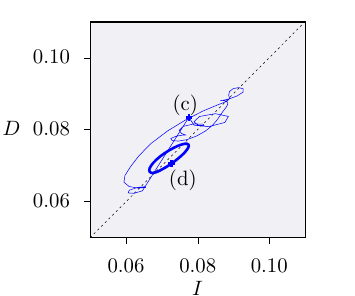}
\caption{Projections of the guess (thin) and converged (thick) loops for the two examples discussed. The energy input is defined as $I=8\int u\sin{4y}\mathrm{d}x\mathrm{d}y/\pi^2 Re$, and the dissipation as $D=8\int \left(\left|\nabla u\right|^2+\left|\nabla v\right|^2\right)\mathrm{d}x\mathrm{d}y/\pi^2 Re^2$, both of which have been normalised by the laminar solution \citep[see][]{chandler2013invariant}.
The labels correspond to the snapshots shown in figure \ref{fig:slices}.}
\label{fig:projection}
\end{figure}

\begin{figure}
    \centering
    \setlength{\fboxsep}{0pt}
    \setlength{\fboxrule}{0.5pt}
    \fbox{\begin{overpic}[width=3cm,angle=90,origin=c]{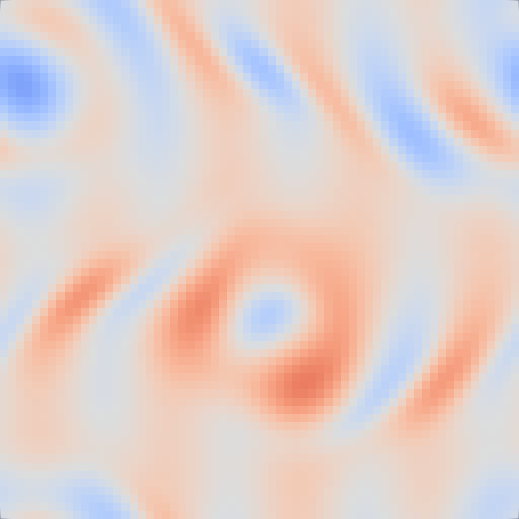}
    \put (10,80) {(a)}
    \end{overpic}}
    \fbox{\begin{overpic}[width=3cm,angle=90,origin=c]{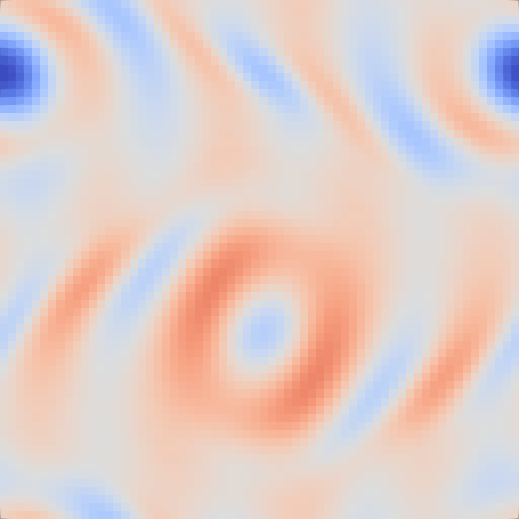}
    \put (10,80) {(b)}
    \end{overpic}}
    \fbox{\begin{overpic}[width=3cm,angle=90,origin=c]{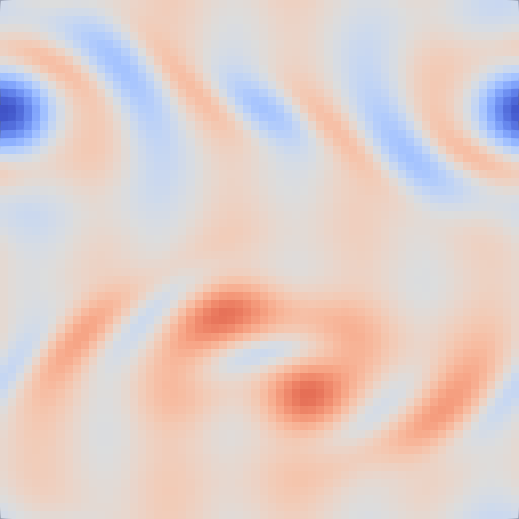}
    \put (10,80) {(c)}
    \end{overpic}}
    \fbox{\begin{overpic}[width=3cm,angle=90,origin=c]{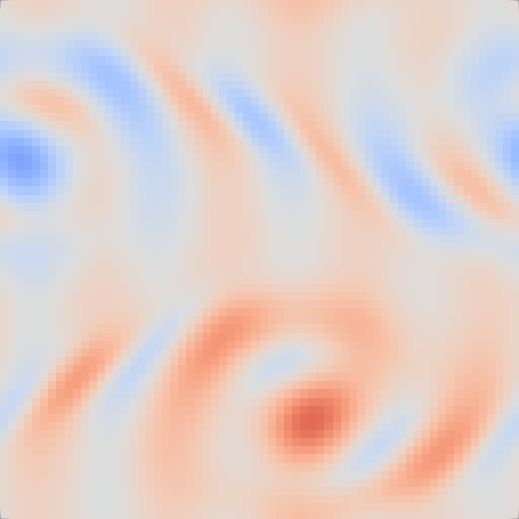}
    \put (10,80) {(d)}
    \end{overpic}}
    \caption{Slices at $s=0$ of the vorticity field for the guess (a,c) and converged (b,d) loops discussed in the text, at the snapshots labelled in figure \ref{fig:projection}. Movies are available in the supplementary materials.}
    \label{fig:slices}
\end{figure}

To compare the performance of the SV, PV and PV-LP methods, we discuss the convergence starting from two initial candidates, which were found through a recurrent flow analysis (see section \ref{sec:rfa}).
Figure \ref{fig:projection} depicts a phase space projection of these two candidates as well as the converged solutions, and figure \ref{fig:slices} shows the flow structure at single snapshots in time.
The first candidate converged to a simple periodic orbit of period $T=5.38$, given as `P1' in \citet{chandler2013invariant}, and the second to the relative periodic orbit `R19' with $T=12.2$, although the algorithm has converged to two periods of this orbit, to give $T=24.4$.
Note how vastly more complicated the initial guess is than the converged solution in this latter case, at least in this simple projection, hinting at the ability of the variational methods to converge to solutions from very distant initial guesses.
The initial candidates were taken from a time-series, with the 64 temporal collocation points for each loop evenly distributed along the series between the starting point and the nearest recurrence. The loop is then closed, which gives the appearance of a sharp discontinuity visible in figure \ref{fig:projection}. Since temporal derivatives are calculated after a Fourier transform, this discontinuity is not problematic.

\begin{figure}
        \centering
        \includegraphics[width=\textwidth]{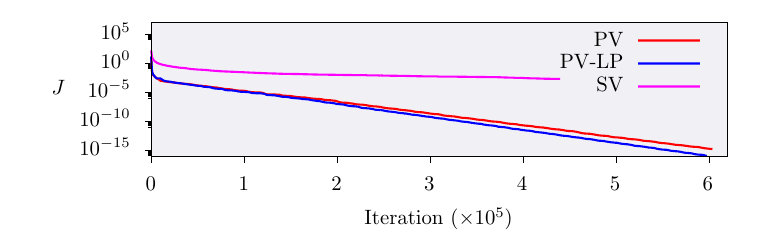}
        \caption{Convergence of the respective residual $J_{PV}$ or $J_{SV}$ for the primitive variables formulation, with (PV-LP) and without (PV) projection and the streamfunction-voriticity (SV) methods applied to the first guess in figure \ref{fig:projection}. In the SV case, $J_{SV}$ starts at a significantly larger value, but also converges much more slowly.}
    \label{fig:convergence}
\end{figure}

Figure \ref{fig:convergence} shows the convergence of the relevant residual against the number of conjugate-gradient iterations, for the first of these two candidates.
The algorithms were each run for three days of wall-clock compute time on a 28 core CPU. Due to the line search requiring more evaluations, each SV iteration is on average around 40 percent slower than the two PV methods, which leads to fewer iterations in 72h. More importantly, SV shows a significantly slower convergence rate. All three methods show an initial fast improvement of the objective function, followed by a much slower period, consistent with \citet{azimi2020adjoint}; the convergence rate in this latter region is much faster with the PV methods, as much smaller steps were required to be taken for SV. The precise reason for this is unclear, and we make no claim that our conjugate-gradient algorithm uses optimal parameters, but this result was robust after significant trial-and-error tweaking of them.
This strongly suggests that the equation \ref{eq:fictitious} of the `fictitious' system is much stiffer for SV than PV.
Whether Leray projection is used has  negligible impact on the rate of convergence, but the path taken towards the solution is certainly different as PV allows for compressible intermediates. Consequently, the methods may converge to different solutions.

\begin{figure}
        \centering
        \includegraphics[width=0.5\textwidth]{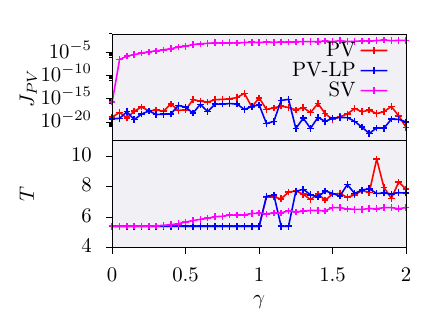}\includegraphics[width=0.5\textwidth]{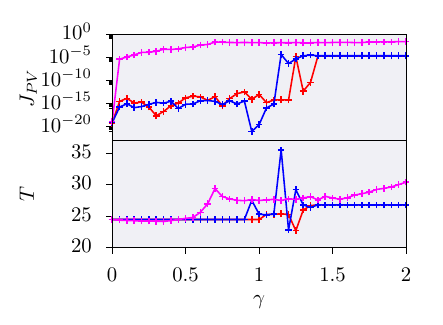}
    \caption{Comparison of methods applied to an initial guess linearly interpolated/extrapolated between the converged periodic orbit ($\gamma=0$) and the recurrent flow analysis candidate ($\gamma=1$). The top plots show the final value of the residual $J_{PV}$, where a value of less than $10^{-10}$ is well converged. The lower plots show final value of the period $T$. Left and right as per the candidates in figure \ref{fig:projection}.}
    \label{fig:comparison}
\end{figure}

To investigate the range of convergence, we systematically vary the initial guess by linearly interpolating between the converged solution and the candidate extracted from flow recurrences, $u_\mathrm{initial} = (1-\gamma) u_\mathrm{solution} + \gamma u_\mathrm{candidate}$, with $\gamma=0$ being the exact (relative) periodic orbit (P1 and R19) and $\gamma=1$ being the candidate. As shown on the top-left in figure \ref{fig:comparison}, for the first candidate,   
the PV formulation, both with and without Leray projection, shows good convergence over the full range of $\gamma$, with $J_{PV}$ consistently reaching less than $10^{-13}$. However, extrapolating to $\gamma>1$, where the initial guess is very different from the periodic orbit P1, the converged period $T$ is shown to rapidly vary. This is because, in this case, the algorithm is no longer converging to P1 but onto a travelling wave solution, for which the period is indeterminate.
Note that around $\gamma=1.2$, the Leray-projected method does converge onto P1 whereas without it does not, which confirms that Leray projection does indeed have a significant role, and that without it, the PV formulation loses the divergence-free property of the guess in the course of converging before again becoming divergence-free as the solution converged, a fact which was confirmed when the divergence was examined.
In contrast with the primitive-variables formulations, the SV formulation performed poorly, with the residual not reaching an acceptably low level within the allotted time for all but the smallest values of $\gamma$, and the final value $T$ being noticeably wrong for $\gamma>0.3$, without the rapid fluctuations indicative of convergence onto a different solution.

For the second, longer period candidate, the results are similar, with convergence good for the primitive variables formulations but much worse for the streamfunction-vorticity formulation. However, in this case, for $\gamma\gtrsim 1.3$, the primitive variables formulations appear to converge to a consistent value of $T$ and a consistent value of $J_{PV}$, which is considerably greater than zero. This, therefore, is a local minimum of the objective functional. This demonstrates the importance of ensuring that the results of the algorithms we find are genuine solutions, rather than just local minima.

\subsection{Comparison with Newton-based shooting}
\label{sec:hybrid}

As the streamfunction-vorticity formulation is signficiantly slower and has worse convergence properties than the primitive variables formulation, in this section we concentrate on the primitive variables formulations, with (PV-LP) and without (PV) Leray projection and compare its performance to a traditional shooting method. 

We compared against a Newton shooting (SN) method. 
This differs significantly from the loop-based methods discussed in this paper. A single point in phase space is integrated in time, using the streamfunction-vorticity formulation of the equations in a comoving frame, up to a guess of the period $T$, with a guess of the phase speed $c$.
The SN residual is then given by the squared distance between the terminal and initial points, defined in our case, due to the implementation, as a slightly unusual metric on the Fourier coefficients of the vorticity.
A Newton-Krylov solver attempts to minimise this so that the trajectory closes back on itself.
Since the handling of time between the methods is very different, a numerically converged solution in SN does not automatically translate to a converged solution in PV, and the converged periods were observed to differ by up to $\pm0.001$. We expect these discrepancies to disappear as the resolutions are increased, though we did not test this.
We used the Newton solver from Channelflow (channelflow.ch) \citep{channelflow}.
With the Hookstep procedure of Channelflow, the algorithm is theoretically globally convergent, but in practice with a poor initial condition the steps can become so small that numerically the algorithm fails to converge, after a few hundred Newton iterations. The method was run for a wall-clock time of 3 days in each case.

\begin{figure}
        \centering
            \includegraphics[width=0.5\textwidth]{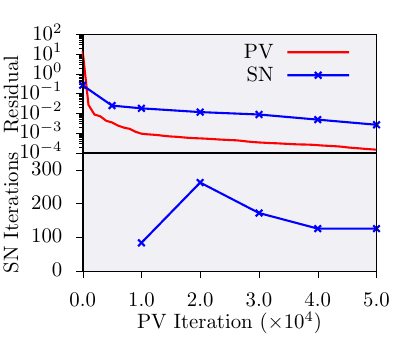}\includegraphics[width=0.5\textwidth]{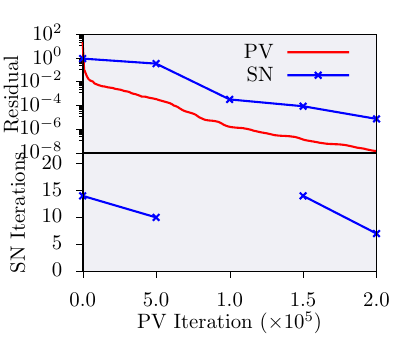}
            \caption{The PV method was paused at intervals and the SN method applied to the result. Top: the SN residual after varying numbers of PV iterations, the (differently defined) residual for which is also shown. Bottom: the number of SN iterations required for convergence to a residual of $10^{-5}$ after the PV method, for those which converged. Where no point is shown, the method did not converge. Left and right as per the candidates in figure \ref{fig:projection}.}
    \label{fig:hybrid}
\end{figure}

It is instructive to apply SN to the two candidates discussed above.
In the first case, the SN procedure described above does not converge starting from the initial guess. In the second case, it does converge, but converges to a simple travelling wave solution (`T1') rather than the relative periodic orbit which the variational methods found. 

We then consider the effects of a partial convergence of the variational methods before applying SN, a hybrid between the two.
The procedure we follow is, after a fixed number of PV iterations, simply to take the first point in the loop as the initial condition for the shooting method, discarding information about all other points on the loop.
Figure \ref{fig:hybrid} shows the results of this on the two candidates discussed before. In both cases, after a number of PV iterations, SN converges rapidly to the same solution as the complete PV method. Once sufficiently many PV iterations have been performed for the result to be within the region of convergence for SN, performing more iterations did not appear to significantly improve the convergence time of SN.
Note the differences between the two solutions: in the first case, many more SN iterations were required, but an order of magnitude fewer PV iterations than in the second case.

The partially converged solutions from PV do not satisfy incompressibility, and so when converting to the streamfunction-vorticity representation used by the SN timestepper (by setting $\omega=\partial_xv-\partial_yu$ and $\psi=-\Delta^{-1}\omega$), some information is lost. This is not the case with the PV-LP formulation, which was found to modestly improve this hybrid approach, with initial SN residuals around one half of the equivalent PV case.

\subsection{Recurrent flow analysis}
\label{sec:rfa}

To test the methods, we implemented a recurrent flow analysis to generate candidates for exact coherent solutions.
Our implementation follows that of \citet{chandler2013invariant}: at regular intervals in a time-series, the current state at time $t$ is compared against previous snapshots at time $t-T$.
The phase space distance, minimised over the phase velocity $c$ in the $x$-direction, is plotted against $t$ and $T$, and local minima below a threshold value are stored as candidate solutions.
We made no attempt to search for relative periodic orbits which exhibit a shift in the $y$ direction, exploiting the discrete symmetry of the equations. There is no reason to suspect that including this would affect the methods we study.
Subsequent authors \citep{lucas2015recurrent, page2020searching} have improved upon this method to find more candidates, but the focus of the present work is merely to investigate the properties of the variational methods. 
A time series of length $200$ was searched every $0.1$ time unit for near-recurrences up to period $T=50$, giving 106 candidate solutions for which the distance $\int \left(\omega(x,y,t)-\omega(x+cT,y,t-T)\right)^2 \mathrm{d}x\mathrm{d}y/\int  \omega(x,y,t)^2 \mathrm{d}x\mathrm{d}y$ was less than the threshold of $0.5$. This is a significantly greater threshold value than that used by \citet{chandler2013invariant}.

Table \ref{table} summarises the results from trying to converge each of these candidates via PV, PV-LP and SN.
We counted as converged any solution for which the residual $J_{PV}$ was less than $10^{-8}$, as the known local minima all had $J_{PV}>10^{-6}$. For all cases except those with particularly long periods, in practice the final value of $J_{PV}$ was significantly smaller than $10^{-8}$.
Each attempted convergence for each of the algorithms was given 72h of wall-clock time on a 28 core CPU, and terminated as soon as the convergence criterion was met, and otherwise counted as a failed convergence.
All three methods were successful at converging solutions, including several that had been found previously and a number of new solutions.
Even with this relatively small sample, it is clear that our new variational methods converge to a greater variety of solutions within the chaotic attractor than the traditional shooting method.
Instances were found where the Newton method converged for a given candidate and the variational methods did not, and vice-versa, hinting at the complex, fractal regions of convergence that are believed to exist for the different methods.

Both our variational methods and the shooting method tend to struggle more in converging orbits with longer periods compared with short ones, though the reasons for this differ.
As the shooting method time-marches the orbit, exponential error amplification leads to sensitive dependence of the recurrence condition on initial conditions and thus an ill-conditioned root search. 
It is for this reason that multi-shooting methods have been developed \citep{sanchez2010multiple}.
In our variational methods, the loop structure means that recurrence is `built in'. Nevertheless we observe that the algorithm converges more slowly for longer orbits. 
Indeed, a large number of candidates appeared to be converging to long-period orbits when the algorithms were terminated and may well have converged had they been allowed to continue. 
In other cases, definite local minima were found by the variational methods, at which point convergence is no longer possible. 
However, switching to a different variational method (e.g. from PV to PV-LP) at this point was found to subsequently lead to convergence to a solution.

\begin{table}
    \centering
\begin{tabular}{c c c c c c}
     $c$ & $T$ & Name & \multicolumn{3}{c}{Number converged} \\
     & & & PV & PV-LP & SN \\\hline
     0 & - & E1 & 7 & 7 & 7  \\
     0.0198 & - & T1 & 26 & 24 & 22 \\
     0 & 5.38 & P1 & 9 & 11 & 3 \\
     0 & 2.83 & P2 & 2 & 4 & 1 \\
     0 & 2.92 & P3 & 2 & 0 & 0 \\
     0.0352 & 12.2 & R19 & 2 & 2 & 0 \\
     0.0173 & 36.8 & R47 & 2 & 3 & 0 \\
     0.00446 & 7.16 & - & 0 & 0 & 1 \\
     0.0213 & 18.1 & - & 0 & 0 & 2 \\
     0.00243 & 3.78 & - & 1 & 0 & 0 \\
     0.0472 & 8.46 & - & 1 & 0 & 0 \\
     0.00780 & 8.64 & - & 1 & 0 & 0 \\
     0.00352 & 9.62 & - & 1 & 2 & 0 \\
     0.0106 & 12.5 & - & 1 & 1 & 0 \\
     0.00993 & 15.7 & - & 1 & 1 & 0 \\
     0.0208 & 18.6 & - & 1 & 0 & 0 \\
     \hline
     \multicolumn{3}{c}{Total} & 57 & 55 & 36 \\
     \multicolumn{3}{c}{Unique} & 14 & 9 & 6
\end{tabular}
\caption{Results of different methods applied to candidates from the recurrent flow analysis. The names are from \citet{chandler2013invariant}, for those solutions which are not new.
All values are given to three significant figures.
Note that the sign of $c$ is irrelevant, due to the symmetries of the system.}
\label{table}
\end{table}

\section{Conclusions}
\label{sec:conclusion}

In this paper, we have presented the first use of variational methods to find unstable periodic orbits in the incompressible Navier-Stokes equations. 
Though the algorithms were slow to converge compared with existing techniques, we have demonstrated their apparently larger region of convergence, raising the possibility of converging a larger number and a greater variety of exact solutions in turbulent flows.

We showed that the primitive variables formulation performs notably better than the streamfunction-vorticity formulation, despite the fact that the latter automatically enforces incompressibility and eliminates the pressure variable. Projecting onto the incompressible subspace or including the divergence-free condition in the objective functional makes little practical difference to the algorithm, suggesting that including incompressibility in the cost function but allowing for compressible intermediate fields during minimization is a viable approach.

The natural extension to this work, and an important step to understanding turbulence in practical flows, is to consider three-dimensional flow models with boundaries, such as plane Couette flow or plane Poiseuille flow. This presents a number of issues not considered here. 
Firstly, with a third dimension at moderately high Reynolds number, the number of grid points required would be several orders of magnitude higher. However, all computations in this work were performed on a single CPU with OpenMP parallelisation, and with MPI parallelisation over many CPUs, or indeed GPU parallelisation, it should be possible to apply such methods to much higher dimensional systems, since all operations except the Fourier transforms are local in either spectral or physical space.
Secondly, the presence of boundaries significantly complicates the derivation of gradients of the objective functionals, and care would need to be taken in particular for boundary conditions of the pressure field, which can be a complicated issue \citep{gresho1987pressure}.
Thirdly, the streamfunction-vorticity formulation does not have a simple analogue in three dimensional flow, though since it performed poorly here, this is a minor consideration.

In section \ref{sec:rfa} we showed that our method is able to converge more solutions than a shooting method, but in some cases the shooting method converged when ours did not. This suggests that the best strategy, if the aim is to converge as many different solutions as possible, may be to run both methods simultaneously. Likewise, a hybrid method such as that presented in section \ref{sec:hybrid}, where a variational method is performed for a certain number of iterations, followed by a shooting method, may give the best results in terms of number of different solutions converged in a given time, though the number of iterations required is likely to depend strongly on the system in question. In the Kolmogorov flow, many orders of magnitude fewer iterations of the variational method were required to reach a result that the Newton shooting method could converge, versus achieving full convergence with the variational method. Due to the slow convergence of the variational methods, given a sufficiently diverse set of good guesses of periodic orbits, the existing Newton shooting methods may still be the most efficient use of compute time. However, in situations where initial guesses are insufficiently accurate for the Newton method to converge, variational methods will increase the success rate.

All the solutions presented in this work were converged from near-recurrences of the state in a time-series. This simple procedure has been found to be effective, but with the enlarged region of convergence that our method provides, other ways of finding initial guesses are likely to be useful. For example, \citet{page2020searching} used dynamic mode decomposition to find unstable periodic orbits of much longer duration, in which cases near-recurrence becomes unlikely.
This becomes more relevant at higher $Re$ than we have considered here, where near-recurrences are harder to observe as periodic orbits become less stable.

The ultimate goal of this field is to be able to use periodic orbit theory predictively in fully developed turbulence, for which it will be necessary to find a very large number of diverse periodic orbits. The methods presented here are a step in that direction, though it is clear that without new advances in efficiency, they will have to be used in conjunction with Newton-based methods where good initial guesses are available.

\section*{Acknowledgements}
This work was supported by 
the European Research Council (ERC) under the European
Union’s Horizon 2020 research and innovation programme
(Grant No. 865677). 
The authors wish to thank O. Ashtari and S. Azimi for many fruitful discussions.

\section*{Declaration of interests}
The authors report no conflict of interest.

\bibliographystyle{jfm}
\bibliography{references}

\end{document}